\def\leti{Lense--Thirring}

\def\rfr#1{(\ref{#1})}

\def\leti{Lense--Thirring}

\def\dert#1#2{\frac{{{d}}{#1}}{{{d}}{#2}}}

\def\eqi{\begin{equation}}
\def\eqf{\end{equation}}
\def\eqia{\begin{eqnarray}}
\def\eqfa{\end{eqnarray}}

\def\rp#1#2{{#1\over#2}}

\def\ct#1{\cite{#1}}
\def\lb#1{\label{#1}}

\documentclass[11pt]{article}
\usepackage{amsmath,amsthm,amscd,amssymb}
\usepackage{latexsym}
\usepackage{graphicx,epsfig}
\usepackage{threeparttable}

\begin{document}

\noindent{\bf \LARGE{How to reach a few percent level in
determining the Lense-Thirring effect?}}
\\
\\
{Lorenzo Iorio}\\
{\it Dipartimento Interateneo di Fisica dell' Universit${\rm
\grave{a}}$ di Bari
\\Via Amendola 173, 70126\\Bari, Italy}
\\
\\
Eelco Doornbos\\{\it Faculty of Aerospace Engineering, Delft
University Of
Technology,\\
Kluyverweg 1, 2629 HS,\\ Delft, The Netherlands}

\begin{abstract}
In this paper we explore the possibility of suitably combining the
nodes $\Omega$ of the existing geodetic LAGEOS, LAGEOS II and
Ajisai laser-ranged satellites and of the radar altimeter Jason--1
satellite in order to increase the accuracy in testing the general
relativistic gravitomagnetic Lense--Thirring secular effect in the
gravitational field of the Earth. The proposal of introducing
Ajisai and Jason--1 in such a combination comes from the expected
benefits which could be obtained in reducing the aliasing secular
impact of the classical part of the terrestrial gravitational
field. According to the recently released EIGEN-CG01C combined
GRACE+CHAMP+terrestrial gravimetry/altimetry Earth gravity model,
the impact of the static part of the mismodelled even zonal
harmonics of geopotential, which represent the major source of
systematic error, amounts to 1.6$\%$, at 1$-\sigma$ level. It is
better than the error which could be obtained with a two-node
LAGEOS-LAGEOS II only combination (6$\%$ at 1$-\sigma$). Moreover,
the proposed combination would be insensitive also to the secular
variations of the low-degree even zonal harmonics, contrary to the
LAGEOS-LAGEOS II only combination. Such variations could be a
serious limiting factor over observational time spans many years
long. The price to be paid for this improvements of the systematic
error of gravitational origin is represented by the
non--conservative forces introduced along with the new orbital
elements. However, they would induce periodic perturbations,
contrary to the gravitational noise. A major concern would be the
assessment of the impact of the non--conservative accelerations on
the Jason--1 node. According to the present--day force models, the
mismodelling in the non--conservative forces would, at worst,
induce an aliasing periodic signal with an amplitude of 4$\%$ of
the Lense-Thirring effect over a time span of 2 years. However, an
observational time span of just some years could safely be adopted
in order to fit and remove the residual long--period
non--gravitational signals affecting Jason's node, which, in the
case of the direct solar radiation pressure, have a main
periodicity of approximately 120 days. Of course, the possibility
of getting time series of the Jason's node some years long should
be demonstrated in reality.
\end{abstract}

\section{Introduction} \subsection{The performed attempts to measure
the Lense--Thirring effect with the LAGEOS satellites}

The general relativistic gravitomagnetic force \ct{ciu95}, related
to the component of the gravitational field induced by the
rotation of a central body of mass $M$ and proper angular momentum
$J$, is still awaiting for a direct, unquestionable measurement.
Up to now there exist some indirect evidences of its existence as
predicted by the General Theory of Relativity (GTR in the
following) in an astrophysical, strong-field context \ct{ste03}
and, in the weak-field and slow-motion approximation valid
throughout the Solar System, in the fitting of the ranging data to
the orbit of Moon with the Lunar Laser Ranging (LLR) technique
\ct{nor03}.

The direct measurement of the gravitomagnetic Schiff precession of
the spins of four spaceborne gyroscopes \ct{sch60} in the
gravitational field of Earth is the goal of the Stanford GP-B
mission \ct{eve01} which has been launched in April 2004. The
claimed obtainable accuracy is of the order of 1$\%$ or better.

\begin{figure}
\begin{center}
\includegraphics{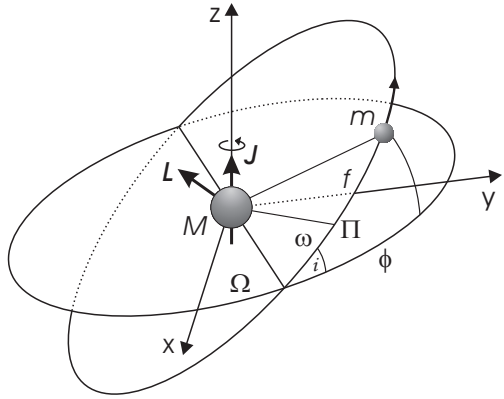}\label{orbita}
\end{center}
\caption{\label{figura1}Orbital geometry for a motion around a
central mass. Here $L$ denotes the orbital angular momentum of the
particle of mass $m$, $J$ is the proper angular momentum of the
central mass $M$, $\Pi$ denotes the pericentre position, $f$ is
the true anomaly of $m$, which is counted from $\Pi$, $\Omega$,
$\omega$ and $i$ are the longitude of the ascending node, the
argument of pericentre and the inclination of the orbit with
respect to the inertial frame $\{x,y,z\}$ and the azimuthal angle
$\phi$ is the right ascension counted from the $x$ axis. When
orbits with small inclinations are considered, the longitude of
pericentre $\varpi=\Omega+\omega$ is used. }
\end{figure}

Another consequence of the gravitomagnetic field is the
Lense-Thirring effect on the geodesic path of a test particle
freely orbiting a central rotating body \ct{leti18}. It consists
of tiny secular precessions of the longitude of the ascending node
$\Omega$ and the argument of pericentre $\omega$ of the orbit of
the test particle
\begin{equation}
\dot\Omega_{\rm LT} =\frac{2GJ}{c^2 a^3(1-e^2)^{\frac{3}{2}}},\
\dot\omega_{\rm LT} =-\frac{6GJ\cos i}{c^2
a^3(1-e^2)^{\frac{3}{2}}},
\end{equation} where $a,\ e$ and $i$ are the semimajor axis, the eccentricity and the inclination, respectively, of
the orbit, $c$ is the speed of light and $G$ is the Newtonian
gravitational constant. See Figure \ref{orbita} for the orbital
geometry.
The LAGEOS III/LARES mission \ct{ciu86, iorlucciu02} was
specifically designed in order to measure such effect, but, up to
now, in spite of its scientific validity and relatively low cost,
it has not yet been approved by any space agency or scientific
institution. Recently, a drag-free version of this project
\ct{ioretal04}, in the context of the relativistic OPTIS mission
\ct{lam01}, is currently under examination by the German Space
Agency (DLR).
\section{The current LAGEOS-LAGEOS II Lense-Thirring experiment}
The Ciufolini's team analyzed the laser-ranged data to the
existing geodetic LAGEOS and LAGEOS II satellites over time spans
of some years \ct{ciu98, ciu02}  by using the following
$J_2-J_4-$free\footnote{$J_2$ and $J_4$ are the first two even
zonal harmonic coefficients of the multipolar expansion of the
terrestrial gravitational field. It turns out that they induce
secular precessions on $\Omega$ and $\omega$. Their mismodelled
parts, according to the currently available Earth gravity models,
are of the same order, or even larger, than the gravitomagnetic
secular precessions of interest.} linear combination of the
orbital residuals of the nodes of LAGEOS and LAGEOS II and the
perigee of LAGEOS II \ct{ciu96} \eqi \delta\dot\Omega^{\rm LAGEOS
}+c_1\delta\dot\Omega^{\rm LAGEOS\ II}+c_2\delta\dot\omega^{\rm
LAGEOS\ II}\sim \mu_{\rm LT}60.2,\lb{ciufform} \eqf where
$c_1=0.304$, $c_2=-0.350$ and $\mu_{\rm LT}$ is the solved--for
least squares parameter which is 0 in Newtonian mechanics and 1 in
the GTR\footnote{It can be expressed in terms of the PPN parameter
$\gamma$ as $\mu_{\rm LT}=\rp{1+\gamma}{2}$. For the PPN formalism
see \ct{wil93, ciu95, wil01}.}. The residuals of the nodal and
perigee rates $\delta\dot\Omega$ and $\delta\dot\omega$ are
determined by comparing these orbital elements at the end of two
orbit computation arcs covering the same time span. These orbit
computations consist of a precise orbit, which has been adjusted
to fit the available tracking data, and a predicted orbit,
starting from an initial state vector which it shares with the
precise orbit, but for which the later states are based on
integration of the classical force models, without adjustment to
the tracking data.  The residuals of the rate of change of the
orbital elements are then calculated by dividing the difference
between these two elements by the time span of the arc. The
signature of the gravitomagnetic force, assumed as an unmodelled
effect, is a linear trend in the combination of the accumulated
residuals, with a slope of 60.2 milliarcseconds per year (mas
yr$^{-1}$ in the following).

The latest, 2002, measurement of the Lense--Thirring effect with
\rfr{ciufform}, obtained by processing the LAGEOS and LAGEOS II
data over a time span of almost 8 years with the orbital processor
GEODYN II of the Goddard Space Flight Center, yields \ct{ciu02}
\eqi\mu_{\rm LT}\sim 1\pm 0.02\pm\delta\mu_{\rm LT}^{\rm
systematic}, \eqf where $\delta\mu_{\rm LT}^{\rm systematic}$
accounts for all the possible systematic errors due to the
mismodelling in the various competing classical forces of
gravitational and non--gravitational origin affecting the motion
of the LAGEOS satellites. In \ct{ciu02} $\delta\mu_{\rm LT}^{\rm
systematic}$ is estimated to be of the order of \eqi\delta\mu_{\rm
LT}^{\rm systematic}=20-30\%.\lb{cazzciuf}\eqf

\subsection{Some possible criticisms to the performed attempts to measure
the Lense--Thirring effect with the LAGEOS satellites} However, it
must be pointed out that several remarks have been made by other
scientists about the estimate \rfr{cazzciuf} \ct{rie02, rie03}.
First of all, it is based on assuming a 13$\%$ systematic error
due to the mismodelling in the even zonal harmonics of
geopotential. This result has been obtained from the full
covariance matrix of the EGM96 Earth gravity model \ct{lem98}  up
to degree $\ell=20$\footnote{In fact, it turns out that the result
does not change if the even zonal harmonics of degree higher than
$\ell=12$ are neglected in the calculation.}. In the EGM96
solution the recovered even zonal harmonics are strongly
reciprocally correlated; it seems, e.g., that this value for the
systematic error due to geopotential is due to a lucky correlation
between $J_6$ and $J_8$ which are not cancelled by \rfr{ciufform}.
The point is that, according to \ct{rie02, rie03}, nothing would
assure that the covariance matrix of EGM96, which is based on a
multi--year average that spans the 1970, 1980 and early 1990
decades, would reflect the true correlations between the even
zonal harmonics during the particular time intervals of a few
years adopted in the analyses by Ciufolini and coworkers. Then, a
more conservative, although pessimistic, approach would be to
consider the sum of the absolute values of the errors due to the
single even zonal as representative of the systematic error
induced by our uncertainty in the terrestrial gravitational field
according to EGM96 \ct{ior03, iormor04}. In this case we would get
a pessimistic upper bound of 83$\%$.

Also the use of the perigee of LAGEOS II in \rfr{ciufform} sheds
shadows on the estimate of \rfr{cazzciuf}. Indeed, it turns out
that the perigee of the LAGEOS-like satellites is very sensitive
to a whole host of non--gravitational perturbations of radiative
\ct{luc01} and thermal origin \ct{luc02}, whose impact on the
proposed measurement of the Lense--Thirring effect would be very
difficult to reliably assess \ct{rie02, rie03}.  It should also be
pointed out that at the time of the analyses by Ciufolini and
coworkers, non--gravitational perturbations of thermal origin,
such as the solar Yarkovsky--Schach effect, were not present at
all in the force models of the orbital determination software
GEODYN II, which was used in their analyses.

\section{The new perspectives opened up by the CHAMP and GRACE models}
The recent and forthcoming improvement in our knowledge of the
Earth gravity field thanks to CHAMP \ct{pav00} and especially
GRACE \ct{rie02}, opens new possibilities to obtain a more
accurate and reliable measurement of the Lense--Thirring effect
with the LAGEOS satellites as well as the other currently existing
laser--tracked satellites. In the pre--CHAMP and GRACE era these
other satellites would have been unsuitable \ct{ior02} because of
their sensitivity to the higher degree even zonal harmonics of
geopotential, due to their smaller altitude with respect to LAGEOS
and LAGEOS II.  In \ct{iormor04} a new observable has been
explicitly put forth, which consists of a $J_2-$free combination
of the nodes of LAGEOS and LAGEOS II only. It is\footnote{The
possibility of using only the nodes of the LAGEOS satellites in
view of the more accurate GRACE Earth gravity solutions was
presented for the first time in \ct{rie02}, although without
quantitative details.} \eqi\delta\dot\Omega^{\rm LAGEOS}+
q_1\delta\dot\Omega^{\rm LAGEOS\ II }\sim\mu_{\rm
LT}47.9,\lb{iorform}\eqf with $q_1=0.546$. According to the
recently released EIGEN-CG01C Earth gravity model\footnote{Such
model represents a long-term averaged solution which combines data
from CHAMP (860 days), GRACE (109 days) and surface
gravimetry/altimetry; moreover, the released sigmas of the
spherical harmonic coefficients of the geopotential are not the
mere formal statistical errors, but are calibrated, although
preliminarily. Then, guesses of the impact of the systematic error
due to the geopotential on the measurement of the Lense-Thirring
effect based on this solution should be rather realistic. However,
caution is advised in considering the so obtained evaluations
because of the uncertainties of the calibration process which
affect especially the even zonal coefficients \ct{rie03}.}
\ct{rei04a}, the systematic error due to the remaining uncancelled
even zonal harmonics $J_4,\ J_6,\ J_8...$ amounts to 5$\%$. This
result has been obtained with a 1-$\sigma$ RSS (Root Sum Square)
calculation based on the variance matrix of EIGEN-CG01C because,
contrary to EGM96 and the other previous Earth gravity models, in
this case the recovered multipolar coefficients are well
disentangled. A pessimistic 1-$\sigma$ upper bound of 6$\%$ can be
obtained by summing the absolute values of the individual errors.

Note also that the combination \rfr{iorform} preserves one of the
most important features of the combination  \rfr{ciufform} of
orbital residuals: indeed, it allows to cancel out the very
insidious 18.6-year tidal perturbation which is a $\ell=2,\ m=0$
constituent with a period of 18.6 years due to the Moon's node and
nominal amplitudes of the order of 10$^3$ mas on the nodes of
LAGEOS and LAGEOS II \ct{ior01}. On the other hand, the impact of
the non--gravitational perturbations on the combination
\rfr{iorform} over a time span of, say, 7 years could be
quantified in 0.1 mas yr$^{-1}$, yielding a 0.3$\%$ percent error.
The results of Table 2 and Table 3 in \ct{iorlucciu02} have been
used. It is also important to notice that, thanks to the fact that
the periods of many gravitational and non--gravitational
time--dependent perturbations acting on the nodes of the LAGEOS
satellites are rather short, a reanalysis of the LAGEOS and LAGEOS
II data over not too many years could be performed. As already
pointed out, this is not so for the combination \rfr{ciufform}
because some of the gravitational \ct{ior01} and
non--gravitational \ct{luc01} perturbations affecting the perigee
of LAGEOS II have periods of many years.

A possible weak point of \rfr{iorform} could be represented by the
fact that it is affected by the secular variations of the
uncancelled even zonal harmonics $\dot J_4,\ \dot J_6,...$. For
the topic of $\dot J_{\ell}$, which has recently received great
attention by the geodesists' community in view of unexpected
variations of\footnote{Fortunately, any issues concerning $\dot
J_2$ do not affect the combination \rfr{iorform}.} $J_2$, see
\ct{cox02}. The mismodelled shift of \rfr{iorform} due to the
secular variations of the uncancelled even zonal harmonics can be
written as \eqi \sum_{\ell=2}\left(\dot\Omega_{.\ell}^{\rm
LAGEOS}+q_1\ \dot\Omega_{.\ell}^{\rm LAGEOS\ II
}\right)\frac{\delta\dot J_{\ell}}{2}T^2_{\rm obs},\eqf where the
coefficients $\dot\Omega_{.\ell}$ are $\partial \dot\Omega_{\rm
class}/\partial J_{\ell}$ and have been explicitly calculated up
to degree $\ell=20$ in \ct{ior03}. It must be divided by the
gravitomagnetic shift of \rfr{iorform} over the same observational
time span \eqi \left(\dot\Omega_{\rm LT}^{\rm LAGEOS}+q_1\
\dot\Omega_{\rm LT}^{\rm LAGEOS\ II} \right) T_{\rm obs}=47.9\
{\rm mas\ yr^{-1}}\ T_{\rm obs}.\eqf
 By assuming $\delta\dot J_4=0.6\times 10^{-11}$
yr$^{-1}$ and $\delta\dot J_6=0.5\times 10^{-11}$ yr$^{-1}$ (Cox
$et\ al$ 2002), it turns out that the 1-$\sigma$ percent error on
the combination \rfr{iorform}, which grows linearly with $T_{\rm
obs}$, would amount to $1\%$ over one year. However, it must be
pointed out that it is very difficult to have reliable evaluations
of the secular variations of the higher degree even zonal
harmonics of geopotential also because very long time series from
the various existing laser-ranged targets are required.

Recently, \rfr{iorform} has been adopted for a 11-years analysis
\ct{ciu04} based on the 2nd generation GRACE-only EIGEN-GRACE02S
Earth gravity model \ct{rei04b}, although the claimed total
accuracy of the performed test might be rather optimistic
\ct{ior04}.

Since GRACE is expected to improve especially the mid to high
degree part of the spectrum of the even zonals, in \ct{iormor04}
a $J_2-...J_8-$free combination involving the nodes of the
spherical geodetic satellites LAGEOS, LAGEOS II, Ajisai, Starlette
and Stella\footnote{The current RMS of fit of the computed orbits
to the laser data of these satellites is of the order of 1 cm or
even better, especially for the LAGEOS satellites.} is proposed.
This combination would be affected just by the higher degree
$J_{\ell}$ and would be insensitive to the secular variations of
the lower degree even zonal harmonics. However, due to the low
altitude of Starlette and Stella, the 1-$\sigma$ error due to the
static part of the geopotential would still amount to 10$\%$ (RSS
calculation) with a 1-$\sigma$ upper bound of $31\%$, according to
the present--day EIGEN-CG01C model. It may be interesting to note
that the combination \rfr{ciufform} would be affected less than
1$\%$ by the mismodelling in the static part of geopotential.

\section{The use of the laser-ranged satellite Ajisai and of the radar altimeter satellite Jason-1}
Up to now, the attention has been focused exclusively on the
spherical laser-ranged satellites because of the high accuracy
with which it is possible to determine their orbits. It is so also
thanks to their high altitude, small area--to--mass $S/M$ ratio
and spherical shape, which reduces the impact of the
non--gravitational perturbations. A recent, important achievement
in orbit determination regards Jason--1. It is a radar altimeter
satellite, launched on December 7, 2001, as a follow-on to the
very successful TOPEX/Poseidon mission which was launched in 1992.
The satellite carries state-of-the-art hardware for the three most
accurate tracking systems available: Satellite Laser Ranging
(SLR), Doppler Orbitography and Radiopositioning Integrated by
Satellite (DORIS) and Global Positioning System (GPS). In
addition, its radar altimeter measurements over the oceans can be
used for either independent orbit validation, or as additional
tracking data. In \ct{lut03}  it has been shown that it is
possible to reach the 1--cm level in determining its orbit (in the
radial direction) by using this dense coverage of precise tracking
data.

This is a very important result, because the orbital parameters of
Jason--1\footnote{One could argue that TOPEX/Poseidon, with the
same orbital parameters as Jason--1, a smaller area--to--mass
ratio and a 12--year (and counting) lifetime, would be the more
suitable satellite. However, TOPEX/Poseidon also has a more
irregular attitude and solar array behavior, a more complex shape
and less precise tracking instrumentation. Its RMS radial orbit
accuracy is estimated at 2.8 cm \ct{tap96}. } make it suitable to
be used, in principle, for the purpose of an accurate and reliable
measurement of the Lense--Thirring effect, provided that accurate
force models are available for a dynamic orbit determination.
Indeed, the Jason--1 orbit parameters are rather similar to those
of the geodetic Ajisai satellite. In Table \ref{para} the orbital
parameters of the LAGEOS satellites and of Ajisai and Jason--1 are
reported. However, it must be preliminarily pointed out that
special care should be taken in handling the orbital manoeuvre
burns of Jason--1, which are designed to counteract the drifting
of the orbit and keep it on its repeating ground--track, as well
as the infrequent safe--mode periods, which also complicate the
dynamic modelling. These events do not prevent us from measuring
any long--term drifts in the node however, as long as the orbital
arcs are designed to start after and stop just before these
instances. Because the manoeuvres are mostly performed in pairs,
approximately one hour apart, and the two safe--mode periods up to
now have lasted several days, they would however introduce
uncertainties when fitting for long--period signals.

\begin{table}[ht!] \caption{Orbital parameters, predicted
Lense-Thirring nodal rates $\dot\Omega_{\rm LT}$ and approximate
area--to--mass ratios $S/M$. $a,e,i$ and $n$ are the semimajor
axis, the eccentricity, the inclination to the Earth's equator and
the Keplerian mean motion $n=\sqrt{GM/a^3}$, respectively.
}\label{para}
\begin{center}
\scalebox{0.8}[0.8]{
\begin{threeparttable}
\begin{tabular}{lllllll}
\noalign{\hrule height 1.5pt}

\small{Satellite} & \small{$a$ (km)} & \small{$e$} & \small{$i$
(deg)} & \small{$n$ ($10^{-4}$ s$^{-1}$)} &
\small{$\dot{\Omega}_{\rm LT}$ (mas yr$^{-1}$)}
& \small{$S/M$ (m$^{2}$ kg$^{-1}$)} \\

\hline

\small{LAGEOS}    &  12270    & 0.0045 &  109.84  & 4.643 & 30.7  & \small{6.9 $\times 10^{-4}$}\\
\small{LAGEOS II} &  12163    & 0.0135 & 52.64  & 4.710 &  31.4 & \small{7.0 $\times 10^{-4}$}\\
\small{Ajisai}    & 7870    & 0.001  & 50.0   & 9.042 &   116.2 & \small{5.3 $\times 10^{-3}$}\\
\small{Jason--1}  & 7713    & 0.0001 & 66.04  & 9.320 &   123.4 & \small{2.7 $\times 10^{-2}$}\\

 \noalign{\hrule height 1.5pt}
\end{tabular}
\end{threeparttable}
}
\end{center}
\end{table}

\subsection{A possible combination of nodes and the gravitational errors}
The following combination \eqi\delta\dot\Omega^{\rm
LAGEOS}+k_1\delta\dot\Omega^{\rm LAGEOS\
II}+k_2\delta\dot\Omega^{\rm Ajisai}+k_3\delta\dot\Omega^{\rm
Jason-1}=\mu_{\rm LT}49.5,\lb{jason}\eqf with \eqi k_1=0.347,\
k_2=-0.005,\ k_3=0.068,\lb{jasoncoef}\eqf could be used in order
to cancel the effect of the first three even zonal harmonics
$J_2$, $J_4$ and $J_6$ along with their temporal seasonal,
stochastic and secular variations. Recently, an analogous approach
has been followed \ct{ves04} by including also some GPS
satellites. According to EIGEN-CG01C, the systematic error induced
by the remaining uncancelled even zonal harmonics of geopotential
amounts to 0.7$\%$ only (RSS calculation at 1-$\sigma$ level),
with an upper bound of 1.6$\%$ (sum of the absolute values of the
individual errors at 1-$\sigma$ level). It must be noted that,
contrary to the combination which can be obtained with the nodes
of Stella and Starlette instead of that of Jason--1, a calculation
up to degree $\ell=20$ is quite adequate to assess the error due
to geopotential. It is so thanks to the higher altitude of
Jason--1 with respect to Stella ($a_{\rm Stl}=$ 7193 km) and
Starlette ($a_{\rm Str}=7331$ km). Contrary to the other examined
combinations, in particular the LAGEOS-LAGEOS II two-nodes
$J_2$-free formula of \ct{iormor04}, the secular variations of the
even zonal harmonics $\dot J_{\ell}$ do not affect \rfr{jason}.
The combination \rfr{jason} would not be affected by harmonic
perturbations of tidal origin with large amplitudes and
particularly long periods. Indeed, the tidal perturbation induced
by the tesseral ($m=1$) $K_1$ tide, which is one of the most
powerful tidal constituents in affecting the orbits of Earth
satellites \ct{ior01}, has its period equal to that of the
satellite's node; it amounts to -0.47, -0.32, -1.55 and 2.84 years
for Jason--1, Ajisai, LAGEOS II and LAGEOS, respectively. This
means that a relatively short observational time span could be
adopted in order to have it longer than the periods of the most
effective perturbations which could then be fitted and removed
from the time series \ct{iorpav01, pavior02}.

\subsection{The impact of the observational errors of Ajisai and Jason-1}
In regard to the impact of the measurement errors, let us assume,
in a very pessimistic way, that the RMS of the recovered orbits of
Ajisai and Jason-1, obtained in a truly dynamical way without
resorting to too many empirical accelerations which could absorb
also the Lense-Thirring effect, amount to 1 m over, say, 1 year.
Then, the error in the nodal rates can be quantified as 26.2 mas
and 26.7 mas for Ajisai and Jason-1, respectively. Thanks to the
coefficients \rfr{jasoncoef}, their impact on the combination
\rfr{jason} would amount to 1.6 mas, i.e. 3.4$\%$ of the
Lense-Thirring effect over 1 year.
\subsection{The impact of the non-gravitational perturbations of Ajisai}
Ajisai is a spherical geodetic satellite launched in 1986. It is a
hollow sphere covered with 1436 corner cube reflectors (CCR's) for
SLR and 318 mirrors to reflect sunlight. Its diameter is 2.15 m,
contrary to LAGEOS which has a diameter of 60 cm only. Its mass is
685 kg, while LAGEOS mass is 406 kg. Then, the Ajisai's
area-to-mass ratio $S/M$, which the non--conservative effects are
proportional to, is larger than that of the LAGEOS satellites by
almost one order of magnitude resulting in a higher sensitivity to
surface forces. However, we will show that their impact on the
proposed combination \rfr{jason} should be less than 1$\%$.
\subsubsection{The atmospheric drag}
The most important non--conservative force in affecting the orbits
of the low satellites is the atmospheric drag. Its acceleration
can be written as \eqi
\textbf{\textit{a}}_{D}=-\rp{1}{2}C_D\left(\rp{S}{M}\right)\rho V
\textbf{\textit{V}},\eqf where $C_D$ is a dimensionless drag
coefficient close to 2, $\rho$ is the atmospheric density and
\textbf{\textit{V}} is the velocity of the satellite relative to
the atmosphere (called ambient velocity). Let us write
\textbf{\textit{V}}=\textbf{\textit{v}}$-{\boldsymbol{
\omega}}\times$ \textbf{\textit{r}} where \textbf{\textit{v}} is
the satellite's velocity in an inertial frame. If the atmosphere
corotates with the Earth ${\boldsymbol \sigma}$ is the Earth's
angular velocity vector
${\boldsymbol\omega}_{\oplus}=\omega_{\oplus}$\textbf{\textit{k}},
where \textbf{\textit{k}} is a unit vector. However, it must be
considered that there is a 20$\%$ uncertainty in the corotation of
the Earth's atmosphere at the Ajisai's altitude\footnote{ It is
believed that the atmosphere rotates slightly faster than the
Earth at some
 altitudes with a 10-20$\%$ uncertainty (Ries 2004, private communication).}. We will then
assume ${\boldsymbol
\sigma}=\omega_{\oplus}(1+\xi)$\textbf{\textit{k}}, with $\xi
=0.2$ in order to account for this effect.

Regarding the impact of a perturbing acceleration on the orbital
motion, the Gaussian perturbative equation for the nodal rate is
\eqi\dert{\Omega}{t}=\frac{1}{na\sqrt{1-e^2}\sin
i}A_N\left(\frac{r}{a}\right)\sin u,\lb{nodogaus}\eqf where
$n=\sqrt{GM/a^3}$ is the Keplerian mean motion, $A_N$ is the
out--of--plane component of the perturbing acceleration and
$u=\omega+f$ is the satellite's argument of latitude.

The out--of--plane acceleration induced by the atmospheric drag
can be written as \ct{abd04} \eqi A_N^{(\rm atm
)}=-\rp{1}{2}K_D\sigma\rho vr\sin i\cos u,\lb{atmo}\eqf with \eqi
K_D=C_D\left(\rp{S}{M}\right)\sqrt{k_R}\eqf and \eqi
k_R=1-\rp{2\sigma h\cos i}{v^2}+\left(\rp{\sigma r
\cos\delta}{v}\right)^2 . \eqf The quantities $h$ and $\delta$ are
the orbital angular momentum per unit mass and the satellite's
declination, respectively. By inserting \rfr{atmo} in
\rfr{nodogaus} and evaluating it on an unperturbed Keplerian
ellipse it can be obtained \eqi \rp{d\Omega}{dt}^{(\rm atm)
}\propto -\rp{1}{2}K_D \rho(1-e^2)\sigma a.\lb{atmoden}\eqf It
must be pointed out that the density of the atmosphere $\rho$ has
many irregular and complex variations both in position and time.
It is largely affected by solar activity and by the heating or
cooling of the atmosphere. Moreover, it is not actually
spherically symmetric but tends to be oblate. A very cumbersome
analytic expansion of $\rho$ based on the TD88 model can be found
in \ct{abd04}. In order to get an order of magnitude estimate we
will consider a typical value $\rho=1\times 10^{-18} $ g cm$^{-3}$
at Ajisai altitude \ct{sen96}. By assuming $C_D=2.5$ \rfr{atmoden}
yields a nominal amplitude of 25 mas yr$^{-1}$ for the atmosphere
corotation case and 5 mas yr$^{-1}$ for the 20$\%$ departure from
exact corotation. The impact of such an effect on \rfr{jason}
would be $3\times 10^{-3}$.
\subsubsection{The thermal and radiative forces} The action of the
thermal forces due to the interaction of solar and terrestrial
electromagnetic radiation with the complex physical structure of
Ajisai has been investigated in \ct{sen96}. The temperature
asymmetry on Ajisai caused by the infrared radiation of the Earth
produces a force along the satellite spin axis direction called
Yarkovsky-Rubincam effect. This thermal thrust produces secular
perturbations in the orbital elements, but no long-periodic
perturbations exist if the spin axis of Ajisai is aligned with the
Earth's rotation axis. In fact, the spin axis was set parallel to
the Earth rotation axis at orbit insertion. The analogous solar
heating (Yarkovsky-Schach effect) is smaller than the terrestrial
heating. A nominal secular nodal rate of 15 mas yr$^{-1}$ due to
the Earth heating has been found. It would affect \rfr{jason} at a
$1.5\times 10^{-3}$ level.

The effect of the direct solar radiation pressure on Ajisai has
been studied in \ct{sen95}. For an axially symmetric, but not
spherically symmetric, satellite like Ajisai there is a component
of the radiation pressure acceleration directed along the
sun-satellite direction $a^{(\rm iso)}_{\odot}$ and another
smaller component perpendicular to the sun-satellite direction. By
assuming for the isotropic reflectivity coefficient its maximum
value $C_R=1.035$ the radiation pressure acceleration $a^{(\rm iso
)}_{\odot}$ experienced by Ajisai amounts to $2.5\times 10^{-8}$ m
s$^{-2}$. Its nominal impact on the node, proportional to
$ea^{(\rm iso )}_{\odot}/4na$, can be quantified as 5.7 mas
yr$^{-1}$; it yields a $5.7\times 10^{-4}$ relative error on
\rfr{jason}. The anisotropic component of the acceleration would
amount, at most, to 2$\%$ of the isotropic one, so that its impact
on \rfr{jason} would be totally negligible.
\subsection{The impact of the non-gravitational perturbations on Jason-1}
The complex shape, varying attitude modes and the relatively high
area--to--mass ratio of the satellite (see Table \ref{para}),
suggest a more complex modelling and higher sensitivity to the
non--gravitational accelerations than in the case of the spherical
geodetic satellites. On the other hand, important limiting factors
in the non-gravitational force modelling of the spherical
satellites are attitude and temperature knowledge \ct{luc01}.
These parameters are actually very well--defined and accurately
measured \ct{mar95} on satellites such as Jason--1. A lot of
effort has already been put into the modelling of
non--gravitational accelerations for TOPEX/Poseidon \ct{ant92,
mar94, kub01}, so that similar models \ct{ber02} have been
routinely implemented for Jason--1. These so--called box--wing
models, in which the satellite is represented by eight flat
panels, were developed for adequate accuracy while requiring
minimal computational resources. A recent development is the work
on much more detailed models of satellite geometry, surface
properties, eclipse conditions and the Earth's radiation pressure
environment for use in orbit processing software \ct{doo02, zie03
}. It should be noted that such detailed models were not yet
adopted in the orbit analyses of \ct{lut03}.  In fact, their
results were based on the estimation of many empirical
1-cycle-per-revolution (cpr) along-track and cross-track
acceleration parameters, which absorb all the
mismodelled/unmodelled physical effects, of gravitational and
non-gravitational origin, which induce secular and long-period
changes in the orbital elements. Due to the power of this
reduced--dynamic technique, based on the dense tracking data,
further improvements in the force models become largely irrelevant
for the accuracy of the final orbit. Such improved models remain,
however, of the highest importance for the determination of
$\delta\dot\Omega^{\rm Jason-1}$.

From \rfr{nodogaus} it can be noted that, since we are interested
in the effects averaged over one orbital revolution, the impact of
every acceleration constant over such a timescale would be
averaged out. The major problems come from 1--cpr out--of--plane
accelerations of the form $A_N=S_N\sin u+C_N\cos u$, with $S_N$
and $C_N$ constant over one orbital revolution.

\begin{figure}
\center{\includegraphics[width=1.00\textwidth]{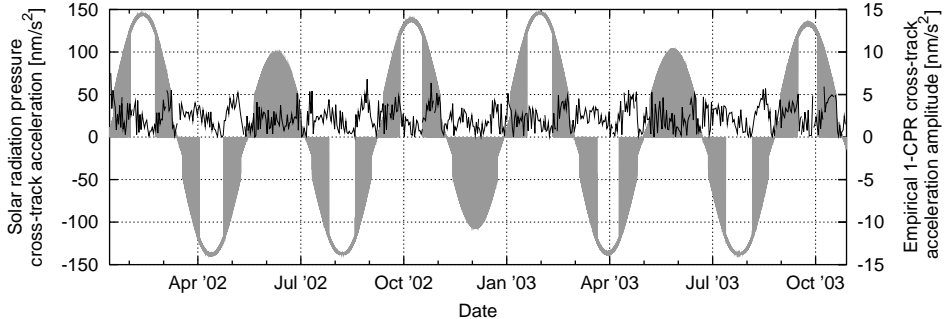}}
\caption{\label{fig:solrad}Time series of the modelled Jason--1
out--of--plane solar radiation pressure acceleration (grey) and
the estimated 1--cpr out--of--plane acceleration (black, right--hand scale).
}
\end{figure}
We have analyzed both the output of the non-gravitational force
models described in \ct{ber02} and the resulting residual
empirical 1--cpr accelerations estimated from DORIS and SLR
tracking over 24--hour intervals. Solar radiation pressure,
plotted in figure \ref{fig:solrad}, is by far the largest
out--of--plane non--gravitational acceleration, with a maximum
amplitude of 147 nm s$^{-2}$. It is followed by Earth radiation
pressure at approximately 7 nm s$^{-2}$.  The contributions of
aerodynamic drag and the thermal imbalance force on the
cross--track component are both estimated to have a maximum of
approximately 0.5 nm s$^{-2}$. As can be seen in figure
\ref{fig:solrad}, the cross--track solar radiation pressure
acceleration shows a sinusoidal long-term behavior, crossing zero
when the Sun--satellite vector is in the orbital plane, roughly
every 60 days. It is modulated by the long--term seasonal
variations in Sun--Earth geometry, as well as by eclipses and the
changing satellite frontal area, both of which are contributing
1--cpr variations. In fact, the shading of certain areas in figure
\ref{fig:solrad} is due to the effect of the eclipses, which, at
once--per--orbit, occur much more frequently than can be resolved
in this figure.

As mentioned before, the empirical 1--cpr accelerations absorb the
errors of almost all the unmodelled or mismodelled forces.  Now
note the systematic way in figure \ref{fig:solrad}, in which the
empirical 1--cpr cross-track acceleration drops to values of below
1 nm~s$^{-2}$ near the end of each eclipse--free period, and has
its maximum level of 5--6 nm~s$^{-2}$ only during periods
containing eclipses.  The fact that the amplitude, but also the
phasing (not shown in the figure) of the 1--cpr accelerations show
a correlation with the orientation of the orbital plane with
respect to the Sun, indicates that it is for a large part
absorbing mismodelled radiation pressure accelerations.

By averaging \rfr{nodogaus} over one orbital revolution and from
the orbital parameters of Table \ref{para} it turns out that a
1--cpr cross--track acceleration would induce a secular rate on
the node of Jason proportional to $7.6 \times 10^{-5}$ s
m$^{-1}\times S_N$ m~s$^{-2}$. This figure must be multiplied by
the combination coefficient $k_3$.  By using the average value of
the empirical 1--cpr acceleration from the above analysis
$S_N\approx$ 2.3 nm~s$^{-2}$ as an estimate for the mismodelled
non--gravitational forces, it can be argued that the impact on
$k_3\delta\dot\Omega^{\rm Jason-1}$ would amount to 77.4 mas
yr$^{-1}$.
However, it must be pointed out that our assumed value of $S_N$
can be improved by adopting the aforementioned more detailed force
models or by tuning the radiation pressure models using tracking
data. In addition, it must be pointed out that $S_N$ experiences
long--term variations mainly induced by the orientation of the
orbital plane with respect to the Sun, and the related variations
in satellite attitude. For Jason--1 such a periodicity amounts to
approximately 120 days (the $\beta^{'}$ cycle). Let us, now,
evaluate what would be the impact of such a long-periodic
perturbation on our proposed measurement of the Lense--Thirring
effect. Let us write, e.g., a sinusoidal law for the
long--periodic component of the weighted nodal rate of Jason--1
\eqi k_3\frac{d\Omega}{dt}=(77.4\ {\rm mas\ yr^{-1}})\times
\cos\left[2\pi\left(\rp{t}{P_{\rm \beta^{'}
}}\right)\right];\lb{booo}\eqf then, if we integrate \rfr{booo}
over a certain time span $T_{\rm obs}$ we get \eqi
k_3\Delta\Omega=\left(\rp{P_{\beta^{'}}}{2\pi}\right)(77.4\ {\rm
mas\ yr^{-1}} )\times\sin\left[2\pi\left(\rp{T_{\rm obs
}}{P_{\beta^{'}}}\right)\right].\eqf Then, the amplitude of the
shift due to the weighted node of Jason--1, by assuming
$P_{\beta^{'}}\cong 120$ days, would amount to \eqi k_3
\Delta\Omega\leq 4\ {\rm mas}.\eqf The maximum value would be
obtained  for \eqi \rp{T_{\rm obs}}{P_{\beta^{'}}}=\rp{j}{4},\
j=1,3,5,...\cong 30, 90, 150,...\ {\rm days}. \eqf So, the impact
on the proposed measurement of the Lense--Thirring effect would
amount to \eqi \left.\rp{\delta\mu_{\rm LT}}{\mu_{\rm
LT}}\right|_{\rm SRP}\leq\rp{(4\ {\rm mas})}{(49.5\ {\rm mas\
yr^{-1}})\times (T_{\rm obs}\ {\rm yr})};\lb{limit}\eqf for, say,
$T_{\rm obs}=2$ years \rfr{limit} yields an upper bound of  $4\%$.
Moreover, it must also be noted that it would be possible to fit
and remove such long--periodic signals from the time series
provided that an observational time span longer than the period of
the perturbation is adopted, as done in \ct{ciu98}.

\section{Conclusions}
According to the EIGEN-CG01C Earth gravity model, which combines
data from CHAMP, GRACE and surface gravimetry/altimetry, the
systematic error in the measurement of the Lense-Thirring secular
effect due to the mismodelling in the static part of the
geopotential is close to a few percent level, at 1-$\sigma$ level,
if the combinations \rfr{jason} is to be used. From this point of
view, \rfr{jason}, which is affected by $J_8, J_{10}...$ at
1.6$\%$ level (1-$\sigma$), is, at present, better than
\rfr{iorform} which, instead, is affected by $J_4, J_6, J_8,...$
at a 6$\%$ level (1-$\sigma$). Another advantage of \rfr{jason}
with respect to \rfr{iorform} is that the combination involving
Jason-1 and Ajisai is insensitive to the secular variations of the
even zonal harmonics, contrary to the two-nodes LAGEOS-LAGEOS II
combination which is, instead, affected by the mismodelling in
$\dot J_4$ and $\dot J_6$. This could represent a drawback because
their effect grows linearly in time and, at present, the knowledge
of $\dot J_{\ell}$ is rather poor. Their estimated impact on
\rfr{iorform} amounts to $1\%$, at 1$-\sigma$ level, over one
year; time spans of more than just one year are required in this
kind of analyses.

The price to be paid for the benefits in the reduction of the
systematic error of gravitational origin is mainly represented by
the non--gravitational perturbations introduced along with the new
orbital elements. However, while the bias due to the geopotential
is unavoidable because it is constant or grows linearly in time,
the signature of the non--conservative accelerations is
long-periodic, so that they could be fitted and removed from the
time series. This perspective should strongly encourage to explore
the possibility of a reanalysis of the Jason--1 orbital data in
order to make them suitable to be combined with those from the
spherical geodetic satellites. Particular attention should be paid
to an as accurate as possible truly dynamical modelling of the
non--gravitational accelerations acting on the node of Jason--1.
This would be helpful also in other scientific branches related to
the Jason-1 activity. It must be noted  the non-gravitational
perturbations, which are almost negligible on the nodes of the
LAGEOS satellites, could represent a major obstacle in using
Jason-1. A reasonable estimate of the impact of the
non--gravitational perturbations modelling error on Jason--1 on
the combination \rfr{jason} yields an error of the order of 4$\%$
over an observational time span of 2 years. However, since the
main periodicity of the direct solar radiation pressure
perturbation seems to amount to approximately 120 days only, over
a time span of a few years it should be possible to average it out
or, at least, to fit and remove such biasing signature from the
time series. Of course, if the combination \rfr{jason} will be
used, a careful choice of the observational time span would be
required in order to reduce the uncertainties related to the
orbital maneuvers, which, however, are mainly-although not
entirely-in plane, and safe--mode events of Jason--1.

Finally,  the obtainable accuracy with the node-node-perigee
combination \rfr{ciufform}, whose error due to geopotential will
remain smaller than that of \rfr{iorform}-\rfr{jason}, is strongly
related to improvements in the evaluation of the
non--gravitational part of the error budget and to the use of time
spans many years long. However, it neither seems plausible that
the error due to the non-conservative forces acting on the perigee
of LAGEOS II will fall to the 1$\%$ level nor that a reliable and
undisputable assessment of it will be easily obtained. Alternative
combinations including the orbital data from the other existing
SLR geodetic spherical satellites are, at present, not competitive
with the combinations \rfr{iorform} and \rfr{jason}; indeed, the
systematic error due to the static part of geopotential amounts to
31$\%$ at 1$-\sigma$ level.

\section*{Acknowledgments}
L.I. thanks H. Lichtenegger for Figure 1.


\end{document}